\documentclass[10pt,journal,compsoc]{IEEEtran}
\usepackage{amsmath,amsfonts}
\usepackage{algorithmic}
\usepackage{algorithm}
\usepackage{array}
\usepackage[caption=false,font=normalsize,labelfont=sf,textfont=sf]{subfig}
\usepackage{textcomp}
\usepackage{stfloats}
\usepackage{url}
\usepackage{verbatim}
\usepackage{graphicx}
\usepackage{threeparttable}
\usepackage{cite}
\usepackage{amsmath,amssymb,amsfonts}
\usepackage{adjustbox}
\usepackage{xcolor}
\usepackage{url,times,booktabs, tabularx}
\usepackage[colorlinks=false,pdfborder={0 0 0}]{hyperref}
\usepackage{units}
\usepackage{multirow,array}
\usepackage{balance}
\usepackage{float}
\usepackage{soul}
\usepackage{color}
\usepackage{cleveref}
\usepackage{wrapfig}
\usepackage{booktabs}
\usepackage[activate]{microtype}
\usepackage{array}
\newcolumntype{P}[1]{>{\centering\arraybackslash}p{#1}}

\newcommand{\eg}{e.\,g.,\ }
\newcommand{\ie}{i.\,e.,\ }

\newcommand{\et}{{et al.\ }}

\newcolumntype{R}[1]{>{\raggedleft\arraybackslash}m{#1}}
\newcommand{\sstitle}[1]{\smallskip\noindent\textbf{#1.\/}}

\begin{document}

\title{
STAA-Net: A Sparse and Transferable Adversarial Attack for Speech Emotion Recognition
}

\author{Yi~Chang,
Zhao~Ren$^{*}$,~\IEEEmembership{Member,~IEEE}, 
Zixing~Zhang,~\IEEEmembership{Senior Member,~IEEE}, 
Xin~Jing, 
Kun~Qian$^{*}$,~\IEEEmembership{Senior Member,~IEEE},
Xi~Shao,~\IEEEmembership{Member,~IEEE},
Bin~Hu,~\IEEEmembership{Fellow,~IEEE}, 
Tanja~Schultz,~\IEEEmembership{Fellow,~IEEE}, Björn~W.~Schuller,~\IEEEmembership{Fellow,~IEEE}

\thanks{This paper was funded partially by the Deutsche Forschungsgemeinschaft (DFG, German Research Foundation) under the project ``MyVoice: Myoelectric Vocal Interaction and Communication Engine'' (460884988), the China Scholarship Council~(CSC), Grant \#\,202006290013, the National Key R\&D Program of China~(No.\,2023YFC2506804), the Ministry of Science and Technology of the People's Republic of China with the STI2030-Major Projects\,2021ZD0201900, the National Natural Science Foundation of China (No.\,62272044), the Teli Young Fellow Program from the Beijing Institute of Technology, China, the National Key Research and Development Project (No.\,2020AAA0106200), and the National Nature Science Foundation of China under Grants (No.\,61936005, No.\,62001038).}
\thanks{Y.\ Chang and B.\,W.\ Schuller are with the GLAM -- the Group on Language, Audio, \& Music, Imperial College London, United Kingdom (y.chang20@imperial.ac.uk, schuller@ieee.org).}
\thanks{Z.\ Ren and T.\ Schultz are with the University of Bremen, Germany (zren@uni-bremen.de, tanja.schultz@uni-bremen.de).}
\thanks{Z.\ Zhang is with the College of Computer Science and Electronic Engineering, Hunan University, China (zixingzhang@hnu.edu.com).}
\thanks{K.\ Qian and B.\ Hu are with the School of Medical Technology, Beijing Institute of Technology, China (qian@bit.edu.cn, bh@bit.edu.cn).}
\thanks{X.\ Shao is with the College of Telecommunications and Information Engineering, Nanjing University of
Posts and Telecommunications, China, (shaoxi@njupt.edu.cn).}
\thanks{X.\ Jing and B.\,W.\ Schuller are also with the Chair of Embedded Intelligence for Health Care and Wellbeing, University of Augsburg, Germany (xin.jing@informatik.uni-augsburg.de).}
\thanks{Corresponding authors: Zhao Ren, Kun Qian}}

\markboth{IEEE Transactions ,~Vol.~XX, No.~XX, XX~2023}%
{Shell \MakeLowercase{\textit{et al.}}: A Sample Article Using IEEEtran.cls for IEEE Journals}


\maketitle

\begin{abstract}
Speech contains rich information on the emotions of humans, and Speech Emotion Recognition (SER) has been an important topic in the area of human-computer interaction. The robustness of SER models is crucial, particularly in privacy-sensitive and reliability-demanding domains like private healthcare. Recently, the vulnerability of deep neural networks in 
the audio domain to adversarial attacks has become a popular area of research.
However, prior works on adversarial attacks in the audio domain primarily rely on iterative gradient-based techniques, which are time-consuming and prone to overfitting the specific threat model. Furthermore, the exploration of sparse perturbations, which have the potential for better stealthiness, remains limited in the audio domain. To address these challenges, we propose a generator-based attack method to generate sparse and transferable adversarial examples to deceive SER models in an end-to-end and efficient manner. We evaluate our method on two widely-used SER datasets, Database of Elicited Mood in Speech (DEMoS) and Interactive Emotional dyadic MOtion CAPture (IEMOCAP), and demonstrate its ability to generate successful sparse adversarial examples in an efficient manner.
Moreover, our generated adversarial examples exhibit model-agnostic transferability, enabling effective adversarial attacks on advanced victim models.
\end{abstract}


\section{Introduction}
\label{sec:intro}
\IEEEPARstart{S}{peech} Emotion Recognition (SER) has been widely applied in many areas in human-computer interaction~\cite{AKCAY202056}, such as diagnosis of depression and
bipolar disorder~\cite{9133298}, computer games~\cite{8925464}, and intelligent call centres~\cite{app122110951}. Recently, due to the rapid development of efficient computing resources and advanced deep learning methods, Deep Neural Networks (DNNs) have achieved good performance in SER~\cite{ZHAO2019312, 8817913, 8462677, 10094895, 10096757, 10141880, 10089511, 10095036}. From the perspective of the input types, there are primarily two categories of DNNs for SER. One category utilises spectrum-based features~\cite{ZHAO2019312, 8817913} while the other processes audio recordings in an end-to-end manner~\cite {8462677, 10095036}. More recently, foundation models (\eg wav2vec 2.0~\cite{NEURIPS2020_92d1e1eb}, or HuBERT~\cite{9585401}) pre-trained on large relative datasets (\eg Librispeech~\cite{panayotov2015librispeech}) have been successfully fine-tuned on smaller dataset for the specific task at hand~\cite{10141880, 10089511}.  

Even though adversarial attacks were first demonstrated and have been extensively studied in the image domain~\cite{advexample2015, madry2018towards, Modas_2019_CVPR, 8601309}, recent research has found that an adversary also poses significant security and privacy threats to the audio domain, such as automatic speech recognition~\cite{carlini2018audio, NeekharaHPDMK19, 3560660, KIM2023109286}, speaker recognition~\cite{9413467, 9519486, 9822974, 9796934}, and classification of acoustic scenes and events~\cite{xie2021enabling}. Moreover, adversarial attacks on SER can also cause severe issues. For instance, these attacks can be used to spread toxic or hateful speech on social media platforms or online gaming platforms, compromising public safety. Additionally, malicious actors can manipulate speech content to deceive automatic speech recognition systems and further manipulate SER models, potentially leading to the propagation of harmful content~\cite{3363271}. Furthermore, in the context of mental illness pre-screening, targeted attacks on speech data could result in incorrect diagnoses and inappropriate treatment for patients~\cite{9133298}. Given these concerns, enhancing the robustness of SER models has emerged as a crucial research area. At present, however, there are relatively few studies~\cite{ren2020enhancing, chang2022robust} on adversarial attacks for SER.

Previous adversarial attack techniques in the audio domain have limitations in terms of their practicality and transferability. White-box approaches (\eg~\cite{carlini2018audio, KIM2023109286, pmlr-v97-qin19a, 9519486}) are often less realistic in the real-world since the adversary has access to all information of the attacked models; iterative gradient-based attacks often suffer from low transferability (\eg~\cite{9155483, JATI2021101199}), which can be attributed to the optimisation of perturbations using the gradient information of the victim model with respect to a specific input. Additionally, most existing attacks on audio tasks impose constraints on the $\boldsymbol{l}_2$~\cite{carlini2018audio, 9822974} or $\boldsymbol{l}_\infty$~\cite{3560660, KIM2023109286, 9413467, 9519486, 9796934} norms of the adversarial perturbations, as the $\boldsymbol{l}_0$ norm is a typical NP-hard problem~\cite{cvpr22_sparseadv, Modas_2019_CVPR}. However, sparse perturbations have the potential to enhance the stealthiness of audio attacks and provide insights into the robustness of DNNs in the audio domain, making $\boldsymbol{l}_0$-constrained attackers worth exploring. Nonetheless, directly adapting existing $\boldsymbol{l}_0$-constrained attackers from the image domain to audio faces two challenges. Firstly, the potential high dimensionality of $1$-D time-sequential audio signals can hinder the efficiency of the attacker. Secondly, the sequential information inherent in audio may not be effectively extracted by such attackers.

To address the aforementioned challenges, we propose STAA-Net, a generator-based adversarial attack method for end-to-end speech SER. Our approach leverages an adjusted Wave-U-Net-like generator to generate sparse audio adversarial perturbations in a single forward pass, enabling efficient and transferable attacks. We validate the effectiveness of STAA-Net through experiments on two widely-used emotional speech datasets. To the best of our knowledge, this study is the first to explore sparse adversarial attacks in the audio domain. Moreover, this work alleviates the scarcity of adversarial attack studies in SER. 

The rest of the paper is organised as follows. Section~\ref{sec:related_work} provides an overview of related studies on SER and adversarial attacks in the audio domain. 
Section~\ref{sec:methods} depicts the detailed methodology. Section~\ref{sec: setup} presents the datasets applied and experimental setups. Section~\ref{sec:results} describes the results and provides an analysis of the findings. Finally, Section~\ref{sec:conclusion} concludes the paper. 

\section{Related Work}
\label{sec:related_work}
\subsection{End-to-end SER}
Even though many SER systems with conventional machine learning techniques utilise hand-crafted acoustic features (\eg Mel-frequency cepstral coefficients (MFCC) features) as input, the selection of these features can introduce bias and take extra time. In recent years, there has been a growing interest in DNNs that directly process raw audio signals, bypassing the need for manual feature extraction and potentially offering more comprehensive representations for SER tasks.

\sstitle{Train-from-scratch Models}
Tzirakis et al.~\cite{8462677} (denoted as Emo18) proposed a convolution recurrent neural network structure initially for continuous emotion recognition (\eg arousal and valence). Emo18 is composed of $3$ convolutional layers for feature extraction from the raw audio signal and $2$-layer Long Short-Term Memory (LSTM) module for contextual dependencies.  
Zhao et al.~\cite{ZHAO2019312} (denoted as Zhao19) proposed a similar network for discrete emotion recognition, and it was composed of $4$ convolutional layers and $2$ stacked LSTM layers. Emo18 and Zhao19 are also utilised and compared in~\cite{TZIRAKIS202146} for the audio modality. 
Zhang et al.~\cite{8682896} employed an attention mechanism and a multi-task learning strategy to enhance the robustness of audio representations for emotion recognition.
Sun et al.~\cite{9170617} utilised a gender information block besides the residual CNN block to improve the recognition accuracy. Tzirakis et al.~\cite{9414866} fused the high-level semantic information from Word2Vec and Speech2Vec models and low-level paralinguistic features extracted by CNN blocks for better performance.

\sstitle{Foundation Models} 
By pre-training on a large amount of data, foundation models learn robust representations that capture both acoustic and linguistic properties of speech, enabling it to transfer knowledge effectively to various speech processing tasks.

Wav2vec 2.0 comprises a Convolutional Neural Network (CNN) module that serves as the feature encoder for latent speech representations, along with a Transformer module that captures global contextual dependencies. Wav2vec 2.0 adopts a self-supervised learning approach, where the model is trained on a massive amount of speech data with a contrastive objective to learn discriminative representations. A Wav2vec 2.0 model~\cite{NEURIPS2020_92d1e1eb} has been widely adopted in SER research~\cite{pepino21_interspeech, 10095036, 10096757, 10089511, cai21b_interspeech} because of its remarkable capability in extracting representations.

The Hidden-Unit BERT (HuBERT)~\cite{9585401} applies an architecture similar to the one of wav2vec 2.0. HuBERT also applies self-supervised learning but with additional auxiliary tasks (\eg frame-wise features predictions), which promotes the model's ability to learn combined acoustic and language features over the raw speech data. Morais et al.~\cite{9747870} fine-tuned HuBERT as upstream model to provide 
generated 
utterance embeddings for emotion classification.

WavLM~\cite{Chen_2022} builds upon the success of self-supervised pre-training in speech processing and aims to tackle full-stack speech processing tasks by leveraging large-scale unlabelled data. To better capture the sequence information in the audio, WavLM further extends the HuBERT approach by employing a gated relative position bias in the Transformer structure and augmenting the training data with an utterance mixing strategy. Feng et al.~\cite{feng2023trustser} applied the WavLM for embeddings extraction and also explored its trustworthiness.

\subsection{Adversarial Attacks in the Audio Domain}
Adversarial attacks in the audio domain can be categorised into two main types: iterative gradient-based attacks and generator-based attacks. Iterative gradient-based attacks typically operate in a white-box setting, utilising gradient information from the victim model to iteratively find minimal perturbations that can deceive the model. Carlini et al.~\cite{carlini2018audio} used a white-box iterative optimisation-based attack to turn any audio waveform into any target transcriptions. Neekhara et al.~\cite{NeekharaHPDMK19} discovered audio-agnostic universal quasi-imperceptible adversarial perturbation through iteratively optimising the normalised Levenshtein distance for automatic speech recognition systems. Kim et al.~\cite{KIM2023109286} found the transferability of adversarial examples is related to the noise sensitivity and proposed a noise injected attack method to generate transferable adversarial examples by iteratively injecting additive noise during the gradient ascent process. Zhang et al.~\cite{9413467} employed a 
Projected Gradient Descent (PGD) attack
with the momentum method to generate text-independent adversarial perturbations for speaker verification systems (SVS). Chen et al.~\cite{9822974} attacked SVS with a 
Fast Gradient Sign Method (FGSM) attack, PGD attack, 
Carlini-Wagner (CW) attack, and FAKEBOB~\cite{9519486} attack to address the optimisation problem.


Recently, generative models such as Generative Adversarial Networks (GAN) \cite{goodfellow2020generative} and autoencoders \cite{vincent2008extracting} have shown promise in generating adversarial perturbations. Compared to iterative gradient-based attackers, generator-based methods focus on learning the distributions of the training data, resulting in more transferable perturbations. Xie et al.~\cite{xie2021enabling} proposed a target attack approach on various audio tasks by concatenating the target class embedding feature map with the intermediate feature map of the generator.

\section{Methods}
\label{sec:methods}
In this section, we first formulate the research problem in Section~\ref{sec:problem}, and then introduce the proposed approach in Section~\ref{sec:proposedapproach}. The description of the loss functions is finally given in Section~\ref{sec:loss}.
\subsection{Problem Formulation}
\label{sec:problem}
We denote an original audio as $\mathbf{x}$, its ground truth emotional class as $y$, added adversarial perturbation as $\boldsymbol{\delta}$, the victim SER model as $\mathit{f}$, and the corresponding adversarial example $\mathbf{x}_{\mathit{adv}} = \mathbf{x} + \boldsymbol{\delta}$. In the un-targeted attack scenario applied in this work, we aim to:

\begin{align*}
\label{eq:target}
\textup{minimise} & \left \| \boldsymbol{\delta} \right \|_0, \\
\textup{subject to } & \textup{argmax}\ \mathit{f}\left ( \mathbf{x}_{\mathit{adv}} \right ) \neq \textup{argmax}\ \mathit{f}\left ( \mathbf{x} \right )\\
\textup{and}\ & \left \| \boldsymbol{\delta} \right \|_\infty  < \epsilon, 
\end{align*}
where $\epsilon$ is a pre-defined hyper-parameter that promotes the imperceptibility of the added adversarial perturbation. However, directly solving the above problem is NP-hard. Inspired by~\cite{10.1007/978-3-030-58542-6_3, cvpr22_sparseadv} in the image domain, we factorise the sparse perturbation $\boldsymbol{\delta}$ to element-wise product of two vectors as follows: 
\begin{equation} 
\label{eq:factor}
\boldsymbol{\delta} = \mathbf{v} \otimes \mathbf{m}, 
\end{equation} 
where $\mathbf{v} \in \mathbb{R}^N$ denotes the perturbation magnitudes, $\mathbf{m} \in \left \{ 0 , 1 \right \}^N$ represents the perturbation locations, $N$ describes the number of time frames of an audio waveform, and $\otimes$ denotes the element-wise product. In the training procedure, $\boldsymbol{l}_1$ regulisation is applied on $\mathbf{m}$ to prompt a sparse perturbation. The above two vectors are optimised separately: one module generates $\mathbf{v}$ and the other module produces $\mathbf{m}$. Because $\mathbf{m}$ is a binary vector, where the value on the time frame $i$ is perturbed if $\mathbf{m}_i = 1$ and unperturbed if $\mathbf{m}_i = 0$, it cannot be directly optimised with gradient back-propagation. To address this issue, a $0$-$1$ random quantisation operation is applied. 
\begin{figure}
    \centering
    \includegraphics[width=\linewidth]{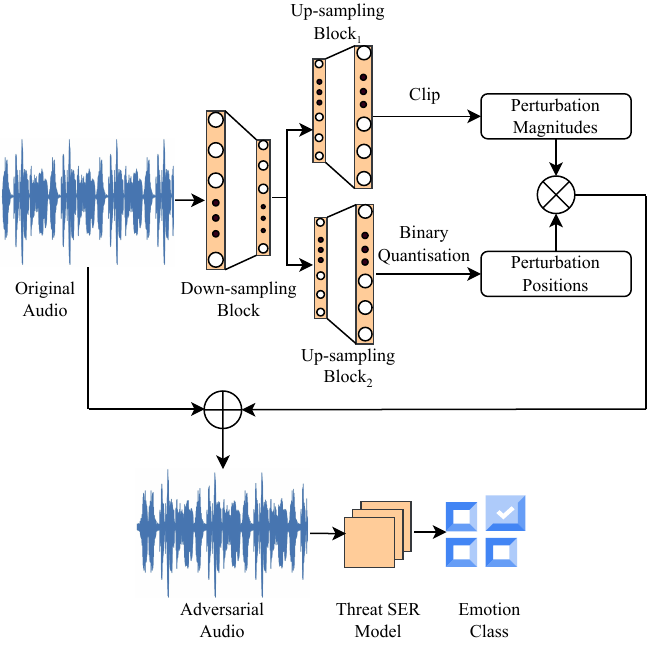}
    \caption{Overall architecture of the proposed STAA-Net.}
    \label{fig:architecture}
\end{figure}

\begin{figure}
    \centering
    \includegraphics[width=0.8\linewidth]{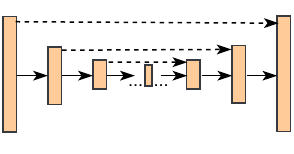}
    \caption{Wave-U-Net architecture. $\rightarrow$ describes down- and up-sampling; $\dashrightarrow$ means concatenation.}
    \label{fig:waveunet}
\end{figure}

\subsection{Proposed Approach}
\label{sec:proposedapproach}
To introduce the proposed approach, the overall architecture is first given in Section~\ref{sec: overall_arc}, followed by the description of the two parts: i) Wave-U-Net and ii) Perturbation Magnitudes and Positions. Finally, we describe the training procedure of the generator.
\subsubsection{Overall Architecture}
\label{sec: overall_arc}
The proposed framework is depicted in Fig.~\ref{fig:architecture}. Since our framework operates on an end-to-end basis, the original raw audio is fed into the generator to produce the perturbation magnitudes and positions. The perturbation magnitudes aim to limit the values of the perturbations at each frame, while the perturbation positions are targeted to be sparse. Next, the sparse perturbations are calculated by multiplying the perturbation magnitudes with the positions. Afterwards, the sparse perturbations are added to the original audio samples to create adversarial audio samples. The training of the generator aims to fool the local threat SER models and the generated adversarial audio samples can also be transferred to attack other unseen targeted SER models.

%
\subsubsection{Wave-U-Net}
U-Net~\cite{unet_2017} was originally developed for accurate and efficient biomedical image segmentation. It utilises a symmetric encoder-decoder structure with skip connections. WaveNet~\cite{vandenoord16_ssw}, on the other hand, is a deep generative model specifically designed for speech synthesis. It leverages the repeated application of dilated convolutions, where the dilation factors increases exponentially, to model the long-term dependencies in audio signals. However, its high memory consumption, attributed to the high sampling rate of audio, has limited its application in real-time scenarios.

Wave-U-Net~\cite{waveunet_2018} extends the U-Net architecture to tackle audio source separation tasks. It combines the benefits of U-Net's encoder-decoder structure and WaveNet's ability to capture long-term dependencies. As depicted in Figure~\ref{fig:waveunet}, in the down-sampling stage, Wave-U-Net reduces the temporal resolution by discarding features for every other time step; in the up-sampling stage, linear interpolation is employed to up-sample the feature maps while also concatenating higher-level features with local features. This approach enables Wave-U-Net to capture longer-term dependencies in audio signals.

As described in Section~\ref{sec:intro}, many previous studies on audio adversarial attacks have predominantly relied on the gradient information of parameters of $\mathit{f}$ with respect to $\mathbf{x}$, leading to limited cross-model transferability. To address these challenges, we propose the adoption of Wave-U-Net~\cite{waveunet_2018} as the generator in this work with one down-sampling block and two up-sampling blocks: $\mathit{UB}_1$ is for perturbation magnitudes and $\mathit{UB}_2$ controls the perturbation locations. 

\subsubsection{Perturbation Magnitudes and Positions}
To generate imperceptible and sparse adversarial audio samples, we divide the final perturbation into magnitudes and positions. On the one hand, the magnitudes control the perturbation values to ensure they are extremely small, thereby making them non-distinguishable from real data. On the other hand, the positions further mitigate the negative impact on audio quality from perturbations by reducing the number of frames affected.

\sstitle{Perturbation Magnitudes}
After $\mathit{UB}_1$, a clip operation is applied to bound the perturbation value into $\{-\epsilon, +\epsilon\}$, where $\epsilon$ is a pre-defined hyper-parameter to constrain the $\boldsymbol{l}_\infty$ norm. The clip operation then leads to the perturbation magnitudes.

\sstitle{Perturbation Positions}
To generate sparse perturbations, we use quantisation to convert the output of $\mathit{UB}_2$ into binary representations as the perturbation positions.
Specifically, if denoting the output of $\mathit{UB}_2$ as $\boldsymbol{\partial}$, in order to transfer it into a discrete vector $\mathbf{m} \in \left \{ 0 , 1 \right \}^N$, we pass $\boldsymbol{\partial}$ into a binary quantisation as: 
\begin{equation}
\label{eq:bi}
    \mathit{q}\left ( \boldsymbol{\partial} _i \right ) = \left\{\begin{matrix}
0, & \boldsymbol{\partial}_i \geq  \gamma & \\ 
1, & \boldsymbol{\partial}_i < \gamma, & 
\end{matrix}\right.
\end{equation}
where $\gamma$ is a hyper-parameter. The binary quantisation described above performs well during inference, but it can encounter issues with gradient vanishing during training~\cite{Gupta_Ajanthan_2022}. To address this, a randomisation approach is introduced to ensure the gradual convergence of $\boldsymbol{\partial}_i$ towards either 1 or 0. Specifically, a random number $t$ is generated from a uniform distribution on the interval $[0, 1)$. If $t \geq 0.5$, the binary quantisation operation is applied; otherwise, the binary quantisation is not performed.

\subsubsection{Training of the Generator}

The training procedure involves feeding the adversarial example $\mathbf{x}_{\mathit{adv}}$ into the threat SER model $\mathit{f}$ for emotion classification. During this process, the generator's parameters are optimised to identify and manipulate significant patterns within the training data. These patterns possess a level of generalisability, which potentially promotes the model-independence of the crafted adversarial examples. 
By leveraging these generalised patterns, the crafted adversarial audio $\mathbf{x}_{\mathit{adv}}$ can effectively deceive and attack other models.

We carefully choose one of the end-to-end train-from-scratch models (\ie Emo18) and one of the pre-trained foundation models (\ie wav2vec 2.0) as local threat model to supervise the generator training. Notably, if the generated sparse adversarial examples, under the guidance of relatively simpler threat model, can still fool more advanced victim models successfully in an efficient manner, it would signify the strong capability of the adversary. Meanwhile, it provides insights into the vulnerabilities and robustness of advanced models when facing adversarial attacks, showcasing the importance of developing defense mechanisms against such attacks.

\subsection{Loss Functions Design}
\label{sec:loss}
The training of the generator is guided by a combination of losses. Specifically, an adversarial loss is used to successfully attack the targeted SER model; a magnitude loss aims to make the adversarial perturbations imperceptible; a sparsity loss encourages sparse perturbations; and a quantisation loss bridges the performance gap between training and inference.

\sstitle{Adversarial Loss}
We employ the Carlini and Wagner (C\&W)~\cite{carlini2017towards} loss as the primary objective function for crafting adversarial audio examples. The C\&W loss aims to increase the probability of misclassification as follows:
\begin{equation}
\mathcal{L}_{adv} = \textup{max}\left \{ \left ( \mathit{f_y}\left ( \mathbf{x}_{\mathit{adv}} \right ) - \underset{i\neq y}{\max} \ \left \{ \mathit{f_i}\left ( \mathbf{x}_{\mathit{adv}} \right ) \right \}\right ), - \phi  \right \},
\end{equation}
where $\phi$ is a confidence pre-defined hyper-parameter to control the attack strength.

\sstitle{Magnitude Loss}
Following the approach of previous works~\cite{3423348, xie2021enabling}, we incorporate an $\boldsymbol{l}_2$ norm regularisation term to ensure that the generated adversarial perturbation remains imperceptible while allowing control over the strength of the attack. This is achieved by adding the $\boldsymbol{l}_2$ norm on the clipped perturbation, as shown in the following equation:
\begin{equation}
\mathcal{L}_{mag} = \left \| \mathbf{v} \right \|_2.   
\end{equation}

\sstitle{Sparsity Loss}
Directly constraining the $\boldsymbol{l}_0$ norm on the added perturbation $\boldsymbol{\delta}$  is NP-hard as discussed beforehand. By factorising the $\boldsymbol{\delta}$ perturbation magnitude $\mathbf{v}$ and perturbation locations $\mathbf{m}$, we can control the sparsity of $\boldsymbol{\delta}$ with the $\boldsymbol{l}_1$ norm of $\mathbf{m}$, since it only contains values $0$ and $1$, where $1$ means the value is perturbed. Therefore, we have the sparsity loss as follows:
\begin{equation}
\mathcal{L}_{spa} = \left \| \mathbf{m} \right \|_1.
\end{equation}
In this way, the sparsity of the final perturbation depends on how the generator converges.

\sstitle{Quantisation Loss}
As explained in Section~\ref{sec: overall_arc}, during the training phase, the decision to perform binary quantisation is determined by a random number sampled from a uniform distribution. In contrast, during inference, binary quantisation is always applied. Consequently, there can be a performance disparity between the generator during training and inference. To mitigate this gap, we introduce the quantisation loss, which is defined as follows:
\begin{equation}
\mathcal{L}_{qua} = \left \| \boldsymbol{\partial} - \mathbf{m} \right \|_2 .   
\end{equation}

\sstitle{Overall Loss}
The overall loss is calculated as the weighted sum of the aforementioned losses, expressed by the equation:
\begin{equation} 
\label{eq:loss_func}
\mathcal{L} = \mathcal{L}_{adv} + \lambda_m \cdot \mathcal{L}_{mag} + \lambda_s \cdot \mathcal{L}_{spa} + \lambda_q \cdot \mathcal{L}_{qua}, 
\end{equation} 
where $\lambda_m$, $\lambda_s$, and $\lambda_q$ represent the weights assigned to the magnitude loss, sparsity loss, and quantisation loss, respectively. These weights allow for fine-tuning the influence of each component in the overall loss function. By optimising the overall loss, the local threat model guides the training of the generator to generate sparse, imperceptible perturbations that achieve a high attack success rate (ASR) (definition can be found in Section~\ref{sec: es}).

\section{Experimental Implementations}
\label{sec: setup}
\subsection{Datasets}

\begin{table}[ht]
\centering
\caption{Emotion distribution of the IEMOCAP dataset.}
\begin{threeparttable}
\begin{tabular}{R{1cm}|R{1cm}|R{1cm}|R{1cm}|R{1cm}|R{1cm}}
\toprule
\# & Anger & Happiness & Neutral & Sadness & $\sum$\\ \hline
Train & \ \ 536 & 1,047 & 1,130 & \ \ 636 & 3,349\\ \hline
Val & \ \ 327 & \ \ 303 & \ \ 258 & \ \ 143 & 1,031\\ \hline
Test & \ \ 240 & \ \ 286 & \ \ 320 & \ \ 305  & 1,151\\ \hline
$\sum$ & 1,103 & 1,636 & 1,708 & 1,084& 5,531\\ \bottomrule
\end{tabular}
\end{threeparttable}
\label{tab:iemocap}
\end{table}

\textbf{DEMoS}: The DEMoS dataset~\cite{parada2019demos} used in this study comprises approximately $7.7$ hours of Italian emotional speech recordings. It involves a total of $68$ speakers, including 23 females and 45 males. Without considering the $332$ neutral speech samples as previous works~\cite{ren2020enhancing, 10096757}, we employ the $9,365$ speech samples (average duration: $2.86$\,seconds $\pm$ standard deviation: $1.26$\,seconds), which are categorised into seven classes: \emph{anger}, \emph{disgust}, \emph{fear}, \emph{guilt}, \emph{happiness}, \emph{sadness}, and \emph{surprise}. We split the dataset into 40\,\% training, 40\,\% validation, 30\,\% testing in a speaker-independent manner. The detailed emotion distribution can be found in our previous work~\cite{10096757}.

\textbf{IEMOCAP}: The Interactive Emotional dyadic MOtion CAPture (IEMOCAP) database~\cite{busso2008iemocap} consists of approximately 12 hours of English audio-visual recordings. Five pairs of actors (a female and a male) participate five recording sessions (\ie 1--5) respectively by either improvising affective scenarios or performing theatrical scripts. The recorded dialogues are further manually segmented into utterances, which are categorised by at least three annotators into different emotional states, \ie \emph{anger}, \emph{disgust}, \emph{excited state}, \emph{fear}, \emph{frustration}, \emph{happiness}, \emph{neutral state}, \emph{sadness}, and \emph{surprise}. 

Similar to prior works on IEMOCAP~\cite{8817913, 9746289, 9761736}, only four classes are included in this work to mitigate the class imbalance, including \emph{anger}, \emph{happiness}, \emph{neutral state}, \emph{sadness}. Moreover, to better compare with prior works~\cite{10095036, 9746289, 9761736}, we merge the class \emph{excited state} into \emph{happiness}. As a result, in total $5,531$ audio samples are applied in this work (average duration: $4.55$\,seconds $\pm$ standard deviation: $3.23$\,seconds). The majority of prior works do not set a validation dataset explicitly and perform $5$-fold cross-validation~\cite{10095036, 9746289, 9761736}. In order to maintain consistency with our experiments on DEMoS and considering that the applied adversarial attack baselines are time-consuming,  we randomly select three sessions (Session $1$, $2$, and $5$) as the training set, one session (Session $4$) as the validation set, and the remaining session (Session $3$) as the test set.
The emotion distribution is described in Table~\ref{tab:iemocap}.

\subsection{Applied Adversarial Attack Baselines}
Many prior works focus on adversarial attacks with $\boldsymbol{l}_2$ or $\boldsymbol{l}_\infty$ constrains, perturbing all values on the time-frame axis, whereas sparse adversarial attacks with $\boldsymbol{l}_0$ constrain target only a few values to fool the victim model. 

\sstitle{Projected Gradient Decent (PGD)}
Projected Gradient Decent is proposed in~\cite{madry2018towards} and it iteratively applies small perturbations to the input data based on the gradient of the loss function with respect to the input. 
\begin{equation} \label{eq:PGD}
x^{i+1} = \mbox{clip}(x^i + \alpha\ \mbox{sign}(\nabla_{x^i}\mathcal{L}(\theta,x^i,y)).
\end{equation}
In each iteration, the magnitude of the perturbation is controlled by a step size $\alpha$ and after each iteration, there is a projection operation $\mbox{clip}(\cdot)$ to make sure the generated current adversarial examples are within a pre-defined range (\ie $\epsilon$-ball). 

Usually, PGD is considered as an $\boldsymbol{l}_2$-constrained attack. However, in this work, to generate sparse adversarial perturbations, we introduce the sparsity constraint directly. Specifically, in each iteration, 
we only perturb a pre-defined number of positions along the time frames of the audio, while keeping the remaining positions unchanged.


\sstitle{SparseFool}
SparseFool proposed in~\cite{Modas_2019_CVPR} is a geometry inspired sparse attack method on an image. By estimating the impact of each pixel on the model's decision boundary with a linear approximation, only the pixels with the highest impact are selected for perturbation. The selected pixels are modified under a sparsity constraint (\ie $\epsilon$) in each iteration.

\sstitle{One-Pixel Attack}
One pixel attack was proposed in~\cite{8601309} with the target to attack models through perturbing just one pixel of the image. Its optimisation-based method has the objective of finding the optimal pixel value and location to lead to the mis-classfication of the attacked model. The optimisation algorithms applied the most are evolutionary strategies. The one-pixel attack also shows some transferability in~\cite{MARRONE2021253}. In this work, we adjust the one-pixel attack method to the 1-D dimension of audio signals and a constraint of the perturbation magnitude $\epsilon$ is also applied for imperceptibility of perturbation.

\subsection{Experimental Settings}
\label{sec: es}
\sstitle{Audio Pre-processing}
All audio recordings are down-sampled to $16$\,kHz for faster processing. For the DEMoS dataset, the audio lengths are unified to the maximum duration of $6.0$ seconds by repeating shorter samples to match the desired length. As for the IEMOCAP dataset, since there is a huge difference between the maximum duration ($34.1$ seconds) and the duration at the $90$-th percentile ($8.7$ seconds), all audio durations are unified to $8.7$ seconds by removing extra signals and repeating shorter samples as necessary.

\sstitle{Models Preparation}
In this work, we experiment with four widely applied end-to-end models with raw audio waves as input: Emo18, Zhao19, wav2vec 2.0, and WavLM. The batch size is $8$ and the model development is supervised by cross-entropy loss. The Emo18 and Zhao19 models are trained from scratch. The training process is optimised with the Adam optimiser with an initial learning rate of $1e-3$ and stopped after $30$ epochs. The wav2vec 2.0 and WavLM models used in this study have been pre-trained on the 960-hour Librispeech corpus~\cite{panayotov2015librispeech}. For the fine-tuning of SER, the feature encoder is frozen, and the classification head consisting of two linear layers is added to make predictions based on the learnt representations. The fine-tuning procedure employs the Adam optimiser with an initial learning rate of $3e-5$, and is stopped after $20$ epochs. 

\sstitle{Implementations of Applied Adversarial Attack Baselines}
The PGD, SparseFool, and one-pixel attack\footnote{https://github.com/Harry24k/adversarial-attacks-pytorch}\footnote{https://github.com/DebangLi/one-pixel-attack-pytorch} methods are adjusted based on their pytorch implementations. For PGD, the number of perturbed locations on the time axis of the audio waveform is set to achieve comparable sparsity with the proposed STAA-Net. The one-pixel attack method specifically sets this number as $1$, while SparseFool dynamically determines the number of perturbed locations. To maintain consistency, a maximum of $20$ iteration steps is set for all three methods. Additionally, to maintain stealthiness as in STAA-Net, the perturbation bound $\epsilon$ is set to $0.05$ for the DEMoS and $0.01$ for IEMOCAP datasets.

\sstitle{Generator Training}
According to their model complexity (number of parameters), we divide the above four models into two sub-groups: Emo18 (1.30\,M) and Zhao19 (1.01\,M), Wav2vec 2.0 (90.37\,M) and WavLM (90.38\,M).
In our study, we carefully select Emo18 and Wav2vec 2.0 as the local threat models to guide the training of our generator. These models provide a solid foundation for supervising the generation of adversarial examples. Additionally, we utilise Zhao19 and WavLM as the victim models to better evaluate the effectiveness of the transferred adversarial examples.

The weights assigned to the magnitude loss $\lambda_m$, sparsity loss $\lambda_s$, and quantisation loss $\lambda_q$ are set as $1e-3$, $1e-4$, and $1e-4$, respectively. As for the binary quantisation, the $\gamma$ is set as $0.5$. When training the generator using end-to-end train-from-scratch models on the IEMOCAP dataset, we observe that the performance of Emo18 is not as good as of the pre-trained models. Consequently, we decide to lower the weights for $\lambda_m$, $\lambda_s$ and $\lambda_q$ by a factor of $0.1$. For the clip operation, we set the bound $\epsilon$ as $0.05$ for DEMoS and $0.01$ for IEMOCAP. This operation ensures that the attack on the IEMOCAP dataset is relatively less potent compared to the attack on the DEMoS dataset.  

The batch size is set to $8$ and we utilise the `Adam' optimiser with an initial learning rate of $1e-4$. The learning rate is decayed by a factor of $0.5$ every $5$ epochs to facilitate convergence. The generator training process is stopped after $20$ epochs.

\sstitle{Evaluations Metrics}
(1) \emph{Unweighted Average Recall (UAR)} is utilised as the standard evaluation metric to mitigate the class imbalance issue~\cite{uar1}, apart from accuracy (\ie weighted average recall). (2) \emph{Attack Success Rate (ASR)} calculates the ratio of the number of adversarial examples causing a miss-classification to the number of total adversarial ones, describing the fooling power of attackers. (3) \emph{Signal-to-Noise Ratio (SNR)} expressed in decibels (dB) measures the relative noise level of perturbation $\boldsymbol{\delta}_i$ to the original audio $\mathbf{x}_i$: $SNR(\mathbf{x}_i, \boldsymbol{\delta}_i) = 20 * \textrm{log}_{10}\frac{max(\mathbf{x}_i)}{max({\boldsymbol{\delta}_i})}$. The larger the SNR, the more imperceptible the added perturbation. According to~\cite{10.1145/3320269.3384733}, an SNR (dB) value close to 20 or larger can be regarded as human imperceptible.

\section{Results and Analysis}
\label{sec:results}
\subsection{SER Models Results}
The Emo18 and Zhao19 models were trained from scratch, while the Wav2vec 2.0 and WavLM models were fine-tuned on the DEMoS dataset. The performance of these models is summarised in Table~\ref{tab:demos_ft}. In comparison to previous works on DEMoS, our applied models demonstrate comparable or even superior performances.
\begin{table}[ht]
\centering
\caption{SER models' performances (Accuracy / UAR [\%]) on DEMoS.}
\begin{threeparttable}
\begin{tabular}{l|R{1.5cm}|R{1.5cm}}
\toprule
Models & Validation & Test \\
\hline
Chang \et~\cite{10096757} & 74.9 / 70.0 & 85.9 / 78.9\\
\hline
Ren \et~\cite{ren2020generating} & -/ 87.5 & -/ 86.7\\
\hline
Ren \et~\cite{10094895} & -/ 91.8 & -/ 91.4\\
\hline
Emo18 & 78.18 / 77.39 & 79.31 / 78.80\\
\hline
Zhao19  & 76.40 / 75.71 & 72.39 / 72.10\\
\hline
Wav2vec 2.0 & \textbf{92.99} / \textbf{92.82} & 91.01 / 91.01\\
\hline
WavLM &  92.48 / 92.30 & \textbf{92.26} / \textbf{92.41}\\
\bottomrule
\end{tabular}
\end{threeparttable}
\label{tab:demos_ft}
\end{table}

For the IEMOCAP dataset, we assess the performance of our applied models in comparison to the state-of-the-art (SOTA) $4$-class emotion recognition methods, as shown in Table~\ref{tab:iemocap_ft}. It is worth noting that the majority of previous works~\cite{10089511} on IEMOCAP employed a dataset split where one session was used as the test dataset and the remaining four sessions were used for training. Some studies (\eg~\cite{santoso_2021, feng2023trustser}) also utilised $5$-fold cross-validation. While our fine-tuned results are comparable to previous works, it is important to consider that differences in performance may arise due to variations in the dataset split. It is important to highlight that this work primarily focuses on the models' performance in the adversarial attack settings, rather than solely on IEMOCAP.
\begin{table}[ht]
\centering
\caption{SER models' performances (Accuracy / UAR [\%]) on IEMOCAP.}
\begin{threeparttable}
\begin{tabular}{l|R{1.6cm}|R{1.6cm}}
\toprule
Models & Validation & Test \\
\hline
Liu \et~\cite{10141880} & - / - & 64.8 / - \\ 
\hline
Pepino \et~\cite{pepino21_interspeech} & - / - & - / 67.2 \\
\hline
Chen \et~\cite{10095036} & - / - & - / 74.3 \\ 
\hline
Santoso \et~\cite{santoso_2021} & - / - & - / \textbf{75.9} \\ 
\hline
Emo18 & 54.03 / 52.81 & 51.26 / 52.40\\
\hline
Zhao19  & 53.93 / 54.17 & 52.48 / 52.74\\
\hline
Wav2vec 2.0 & 67.12 / \textbf{68.07} & 66.46/ 66.74\\
\hline
WavLM & \textbf{68.28} / 67.08 & \textbf{67.07} / 66.90\\
\bottomrule
\end{tabular}
\end{threeparttable}
\label{tab:iemocap_ft}
\end{table}

\subsection{Attack Results}
As shown in Table~\ref{tab:res_demos}, with Emo18 and wav2vec 2.0 as the threat model to train the generator, the produced adversarial examples are used to test on Emo18 and wav2vec 2.0 as white-box attack and also transferred to attack other models in a black-box manner. It needs to be mentioned that SNR value and sparsity are averaged based on the successful adversarial examples for a fair evaluation. Therefore, we can see that adversarial examples generated by the one-pixel attack with Emo18 do not have SNR and sparsity values since the one-pixel attack achieves $0.00$\% attack success rate (ASR). Regarding SNR, all adversarial examples have values around 17 dB, which can be regarded as imperceptible according to~\cite{10.1145/3320269.3384733}. However, the one-pixel attack with wav2vec 2.0 as threat model achieves higher SNR (\ie around $37$ dB), which might be caused by the randomness of just one value on the time frame axis.

In terms of speed, the proposed generator-based method demonstrates rapid generation of adversarial examples (\ie $0.01$ seconds), which is more than $100$ times faster than the PGD attacker when the threat model is Emo18 and about $73$ times faster on the validation dataset and $57$ times faster on the test dataset when the threat model is wav2vec 2.0. Notably, when the attacker is SparseFool and one-pixel, the speed is quite slow, from $8.33$ seconds to $418.36$ seconds to generate one single adversarial examples. Our proposed STAA-Net is quite efficient in this regard. This is essential because in practical attack scenarios, attackers would prefer to quickly generate the adversarial perturbation using mobile devices and inject it into the victim's ongoing speech since attackers usually do not have the opportunities to record and modify the whole speech in real time. This requires a highly efficient method with low computational complexity to craft robust adversarial perturbations within a limited time budget. 

Sparsity and attack performances sometimes are trade-offs, but the proposed STAA-Net can achieve a balance between them. In Table~\ref{tab:res_demos}, when both the threat model and victim model are Emo18, we can see that val ASR $82.71$\% and test ASR $78.50$\% by STAA-Net is considerably
better than those of PGD, SparseFool, and one-pixel with sparser purturbations (\ie val sparsity $12.50$\% and test sparsity $6.25$\%). By limiting the number of perturbation locations to be one, the one-pixel attacker fails to attack any of the audio samples in DEMoS. When attacking other models with Emo18 as the threat model, STAA-Net achieves better transferability. Specifically, the validation and test ASR on Zhao19, wav2vec 2.0, and WavLM are the best, compared with other attackers respectively. Notably, even though the threat model is relatively simple (\ie Emo18), when using generated adversarial examples to attack more advanced models (\ie wav2vec 2.0 and WavLM), the ASR drops but is still quite high. This further proves the effectiveness of the proposed STAA-Net. When the threat model is wav2vec 2.0, if the victim model is also wav2vec 2.0, we can find that ASRs achieved by STAA-Net (\ie val ASR $80.02$\% and test ASR $76.09$\% ) are lower than the ones by PGD (\ie val ASR $98.46$\% and test ASR $99.05$\%). However, the transferability of the STAA-Net is better and the sparsity of the perturbation is quite lower (\ie validation sparsity $11.88$\% $\mathit{vs}$ $22.29$\%  and test sparsity $12.93$\% $\mathit{vs}$ $23.13$\%). Interestingly, when the threat model (\ie wav2vec 2.0) is relatively more advanced than the victim model, the ASR increases a little bit. This might be caused by the strong capability of wav2vec 2.0 for latent representations extraction.

To further validate the effectiveness of STAA-Net, we extend our experiments on the second emotion dataset IEMOCAP and the corresponding results can be found in Table ~\ref{tab:res_iemocap}. In terms of speed, the proposed STAA-Net is the fastest when generating adversarial examples. For SNR, the STAA-Net generates imperceptible adversarial examples, especially when the threat model is wav2vec 2.0. As for the sparsity of the adversarial examples, the STAA-Net-generated adversarial examples are the most sparse, whereas the SparseFool generates quite dense perturbations. When the attack is white-box and the threat model is Emo18, even though the PGD attacker achieves better test ASR (\ie $80.28$\%), the perturbation is denser and transferability is worse. When the threat model is wav2vec 2.0, we can see that adversarial examples generated by STAA-Net have the best transferability and sparsity of perturbations. Sparsefool also fails to generate sparse perturbations.

\begin{figure}
    \centering
    \includegraphics[width=\linewidth]{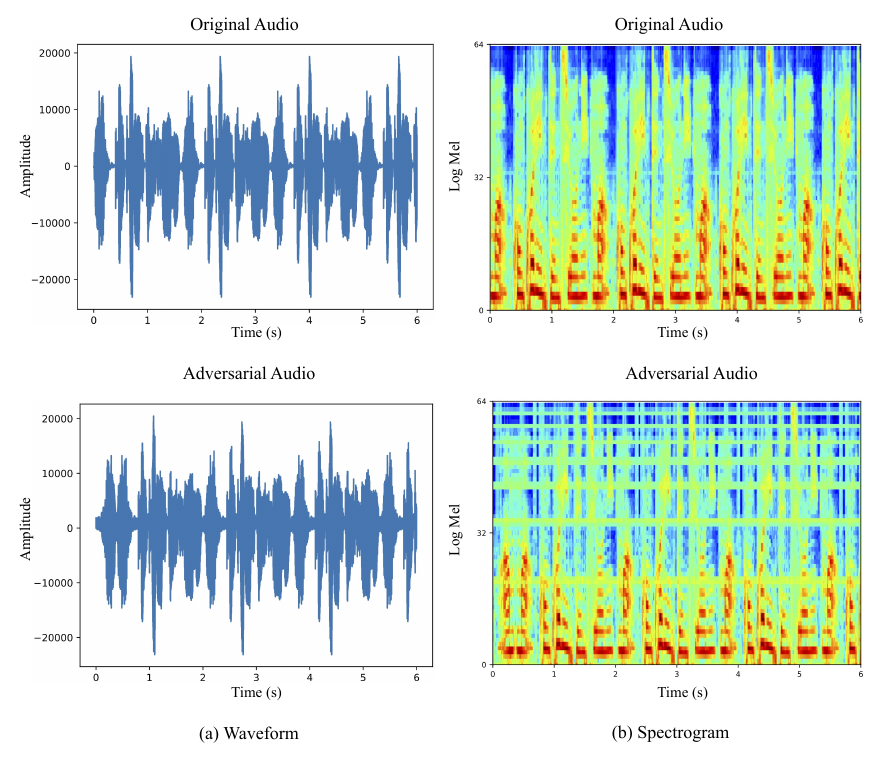}
    \caption{Comparison of the waveforms and log Mel spectrograms of one original audio sample from the DEMoS dataset and its STAA-Net generated adversarial example. The file name of the sample is `NP\_f\_43\_pau05b.wav', spoken by a female with the emotion class `fear'.}
    \label{fig:visalisation}
\end{figure}

\begin{table*}[ht]
    \caption{Attack performances comparison on the DEMoS dataset. The rates [\%] in the columns of Emo18, Zhao19, Wav2vec 2.0, WavLM are Attack Success Rate (ASR).}
    \label{tab:res_demos}
    \centering
    \scalebox{0.85}{
    \begin{threeparttable}
    \begin{tabular}{l|l|R{0.8cm}|R{0.8cm}|R{0.8cm}|R{0.8cm}|R{0.8cm}|R{0.8cm}|R{0.8cm}|R{0.8cm}|R{0.8cm}|R{0.8cm}|R{0.8cm}|R{0.8cm}|R{0.8cm}|R{0.8cm}}
    \toprule
    & &\multicolumn{2}{c|}{\textbf{SNR (dB)}} & \multicolumn{2}{c|}{\textbf{Sparsity (\%)}} & \multicolumn{2}{c|}{\textbf{Speed (s)}} & \multicolumn{2}{c|}{\textbf{Emo18 (\%)}} & \multicolumn{2}{c}{\textbf{Zhao19 (\%)}} & \multicolumn{2}{c}{\textbf{Wav2vec 2.0 (\%)}} & \multicolumn{2}{c}{\textbf{WavLM (\%)}} \\
    \cline{3-16}
    \emph{Threat Model} & \emph{Attacker}& \emph{Val} & \emph{Test} & \emph{Val} & \emph{Test} & \emph{Val} & \emph{Test} & \emph{Val} & \emph{Test} & \emph{Val} & \emph{Test} & \emph{Val} & \emph{Test} & \emph{Val} & \emph{Test}\\

    \hline
    
    & PGD & \textbf{16.87} & 17.38 & 14.58 & 9.94 & 1.10 & 1.16 & 77.85 & 67.78 & 67.09 & 52.65 & 15.50 & 27.66 & 15.87 & 12.60 \\
    & SparseFool & 16.64 & 17.48 & 95.43 & 98.00 & 18.90 & 8.33 & 12.87 & 13.38 & 9.67 & 12.65 & 11.00 & 12.56 & 8.49 & 9.46\\
    Emo18
    & One-Pixel & -- & -- & -- & -- & 29.88 & 24.88 & 0.00 & 0.00 & 0.00 & 0.00 & 0.00 & 0.04 & 0.00 & 0.04\\
    & \textbf{STAA-Net} & 16.86 & \textbf{17.52} & \textbf{12.50} & \textbf{6.25} & \textbf{0.01} & \textbf{0.01} & \textbf{82.71} & \textbf{78.50} & \textbf{84.95} & \textbf{53.63} & \textbf{33.76} & \textbf{41.29} & \textbf{54.31} & \textbf{28.39} \\
    
    \hline
    
    & PGD & 16.87 & 17.53 & 22.29 & 23.13 & 0.73 & 0.57 & 79.93 & 74.19 & 81.11 & 65.12 & \textbf{98.46} & \textbf{99.05} & 48.17 & 33.55 \\
    & SparseFool & 16.69 & 17.31 & 99.61 & 99.73 & 90.64 & 79.87 & 53.40 & 39.31 & 53.88 & 43.23 & 61.65 & 48.47 & 51.01 & 37.03\\
    Wav2vec 2.0
    & One-Pixel & \textbf{36.88} & \textbf{37.54} & 1 (p) & 1 (p) & 82.93 & 71.40 & 0.00 & 0.00 & 0.00 & 0.00 & 0.06 & 0.13 & 0.00 & 0.04\\
    & \textbf{STAA-Net} & 16.87 & 17.21 & \textbf{11.88} & \textbf{12.93} & \textbf{0.01} & \textbf{0.01} & \textbf{82.84} & \textbf{78.07} & \textbf{84.44} & \textbf{82.41} & 80.02 & 76.09 & \textbf{64.67} & \textbf{52.09} \\

    \bottomrule
    \end{tabular}
    \end{threeparttable}
    }
\end{table*}

\begin{table*}[ht]
    \caption{Attack performances comparison on the IEMOCAP dataset. The rates [\%] in the columns of Emo18, Zhao19, Wav2vec 2.0, WavLM are Attack Success Rate (ASR).}
    \label{tab:res_iemocap}
    \centering
    \scalebox{0.85}{
    \begin{threeparttable}
    \begin{tabular}{l|l|R{0.8cm}|R{0.8cm}|R{0.8cm}|R{0.8cm}|R{0.8cm}|R{0.8cm}|R{0.8cm}|R{0.8cm}|R{0.8cm}|R{0.8cm}|R{0.8cm}|R{0.8cm}|R{0.8cm}|R{0.8cm}}
    \toprule
    & &\multicolumn{2}{c|}{\textbf{SNR (dB)}} & \multicolumn{2}{c|}{\textbf{Sparsity (\%)}} & \multicolumn{2}{c|}{\textbf{Speed (s)}} & \multicolumn{2}{c|}{\textbf{Emo18 (\%)}} & \multicolumn{2}{c}{\textbf{Zhao19 (\%)}} & \multicolumn{2}{c}{\textbf{Wav2vec 2.0 (\%)}} & \multicolumn{2}{c}{\textbf{WavLM (\%)}} \\
    \cline{3-16}
    \emph{Threat Model} & \emph{Attacker}& \emph{Val} & \emph{Test} & \emph{Val} & \emph{Test} & \emph{Val} & \emph{Test} & \emph{Val} & \emph{Test} & \emph{Val} & \emph{Test} & \emph{Val} & \emph{Test} & \emph{Val} & \emph{Test}\\

    \hline
    
    & PGD & 23.83 & \textbf{17.04} & 35.96 & 29.19 & 2.20 & 2.26 & 47.14 & \textbf{80.28} & 44.42 & 61.25 & \textbf{37.25} & 41.09 & \textbf{41.64} & 40.40 \\
    & SparseFool & 15.69 & 10.83 & 99.96 & 97.29 & 162.58 & 418.36 & 9.70 & 25.02 & 9.70 & 19.64 & 3.01 & 10.95 & 5.53 & 15.38\\
    Emo18
    & One-Pixel & -- & -- & -- & -- & 40.50 & 69.54 & 0.00 & 0.00 & 0.00 & 0.00 & \, 0.00 & 0.00 & 0.00 & 0.00\\
    & \textbf{STAA-Net} & \textbf{22.99} & 15.99 & \textbf{24.34} & \textbf{24.97} & \textbf{0.01} & \textbf{0.01} & \textbf{58.68} & 71.16 & \textbf{53.15} & \textbf{79.24} & 29.97 & \textbf{46.39} & 36.76 & \textbf{51.17} \\
    
    \hline
    
    & PGD & 26.34 & 19.85 & 8.27 & 15.89 & 0.78 & 0.74 & 17.75 & 54.39 & 19.50 & 47.52 & \textbf{95.64} & \textbf{91.83} & 23.86 & 35.71 \\
    & SparseFool & 27.51 & \textbf{19.87} & 98.41 & 96.32 & 247.38 & 171.09 & 26.77 & 26.59 & 24.64 & 23.46 & 36.47 & 30.67 & 36.86 & 24.41\\
    Wav2vec 2.0
    & One-Pixel & \textbf{40.83} & -- & 1 (p) & -- & 61.64 & 56.00 & 0.00 & 0.00 & 0.00 & 0.00 & 0.19 & 0.00 & 0.19 & 0.00\\
    & \textbf{STAA-Net} & 24.74 & 19.75 & \textbf{3.12} & \textbf{4.61} & \textbf{0.01} & \textbf{0.01} & \textbf{33.95} & \textbf{61.86} & \textbf{27.64} & \textbf{54.56} & 53.35 & 65.33 & \textbf{42.58} & \textbf{45.61} \\

    \bottomrule
    \end{tabular}
    \end{threeparttable}
    }
\end{table*}

\subsection{Ablation Study}

\begin{table*}[ht]
    \caption{Ablation study of STAA-Net on the DEMoS dataset. The rates [\%] in the columns of Emo18, Zhao19, Wav2vec 2.0, WavLM are Attack Success Rate (ASR).}
    \label{tab:res_ablation}
    \centering
    \scalebox{0.8}{
    \begin{threeparttable}
    \begin{tabular}{l|l|R{0.8cm}|R{0.8cm}|R{0.8cm}|R{0.8cm}|R{0.8cm}|R{0.8cm}|R{0.8cm}|R{0.8cm}|R{0.8cm}|R{0.8cm}|R{0.8cm}|R{0.8cm}|R{0.8cm}|R{0.8cm}}
    \toprule
    & &\multicolumn{2}{c|}{\textbf{SNR (dB)}} & \multicolumn{2}{c|}{\textbf{Sparsity (\%)}} & \multicolumn{2}{c|}{\textbf{Speed (s)}} & \multicolumn{2}{c|}{\textbf{Emo18 (\%)}} & \multicolumn{2}{c}{\textbf{Zhao19 (\%)}} & \multicolumn{2}{c}{\textbf{Wav2vec 2.0 (\%)}} & \multicolumn{2}{c}{\textbf{WavLM (\%)}} \\
    \cline{3-16}
    \emph{Threat Model} & \emph{Attacker}& \emph{Val} & \emph{Test} & \emph{Val} & \emph{Test} & \emph{Val} & \emph{Test} & \emph{Val} & \emph{Test} & \emph{Val} & \emph{Test} & \emph{Val} & \emph{Test} & \emph{Val} & \emph{Test}\\

    \hline
    
    & w/o factorisation & 16.92 & 17.60 & 100.00 & 100.00 & 0.01 & 0.01 & 82.99 & 82.41 & 80.33 & 83.83 & 11.70 & 53.29 & 17.35 & 25.16 \\
    & w/o magnitude loss & 16.86 & -inf & 12.50 & 0.00 & 0.01 & 0.01 & 83.26 & 12.86 & 86.34 & 12.77 & 28.50 & 2.37 & 52.01 & 1.59\\
    Emo18
    & w/o sparsity loss & 16.96 & 17.61 & 100.00 & 100.00 & 0.01 & 0.01 & 82.90 & 78.97 & 80.02 & 84.04 & 35.27 & 15.27 & 45.81 & 7.83\\
    & w/o quantisation loss & 16.87 & 17.51 & 12.48 & 6.25 & 0.01 & 0.01 & 83.14 & 78.67 & 85.31 & 64.34 & 29.13 & 41.55 & 51.65 & 28.47\\
    & \textbf{STAA-Net} & 16.86 & 17.52 & 12.50 & 6.25 & 0.01 & 0.01 & 82.71 & 78.50 & 84.95 & 53.63 & 33.76 & 41.29 & 54.31 & 28.39 \\
    
    \hline
    
    & w/o factorisation & 16.88 & 17.43 & 100.00 & 100.00 & 0.01 & 0.01 & 82.77 & 78.32 & 84.38 & 81.03 & 88.03 & 83.91 & 83.05 & 60.86 \\
    & w/o magnitude loss & 16.87 & 17.20 & 22.10 & 18.75 & 0.01 & 0.01 & 82.74 & 78.58 & 85.10 & 74.97 & 83.20 & 72.30 & 70.32 & 52.77\\
    Wav2vec 2.0
    & w/o sparsity loss & 16.91 & 17.40 & 100.00 & 100.00 & 0.01 & 0.01 & 83.35 & 79.91 & 84.47 & 80.00 & 86.82 & 81.72 & 86.37 & 72.90\\
    & w/o quantisation loss & 16.87 & 17.19 & 20.40 & 20.31 & 0.01 & 0.01 & 83.02 & 78.19 & 85.62 & 79.91 & 81.57 & 73.81 & 66.70 & 54.02\\
    & \textbf{STAA-Net} & 16.87 & 17.21 & 11.88 & 12.93 & 0.01 & 0.01 & 82.84 & 78.07 & 84.44 & 82.41 & 80.02 & 76.09 & 64.67 & 52.09 \\

    \bottomrule
    \end{tabular}
    \end{threeparttable}
    }
\end{table*}
We further assess the contributions of the main components within our proposed framework. To this end, we selected Emo18 and wav2vec 2.0 as the local threat models, and the DEMoS dataset as our experimental dataset. The results are indicated in Table~\ref{tab:res_ablation}. The generation of adversarial examples maintains a consistent speed, and there is only little observed alteration in the SNR.

\sstitle{Effects of Factorisation}
In order to assess the effectiveness of the factorisation approach, we remove the up-sampling block $\mathit{UB}_2$ in Figure~\ref{fig:architecture} from our architecture, which controls the perturbation locations, and solely use the output of the other up-sampling block $\mathit{UB}_1$ to generate the final perturbation $\mathbf{v}$. Specifically, 
we retained the magnitude loss $\mathcal{L}_{mag}$, but calculated the sparsity loss as $\mathcal{L}_{spa} = \left \| \mathbf{v} \right \|_1$, while the quantisation loss was omitted. The results, presented in Table~\ref{tab:res_ablation}, demonstrate that the generated adversarial examples achieve $100$\% sparsity.
This outcome suggests that directly applying  $\boldsymbol{l}_1$ regularisation alone does not yield sparse solutions in our study. Furthermore, this ablation study emphasises the significance of $\mathit{UB}_2$ in effectively controlling the locations of the perturbations within the generated adversarial examples.

\sstitle{Effects of Magnitude Loss}
The purpose of using magnitude loss is two-fold: one is to control the attack strength and the second is to alleviate extremely large perturbations. Without the magnitude loss, the generator is not stable and can produce perturbation with all values as $0$ with Emo18 as local threat model on the test dataset of DEMoS as shown in Table~\ref{tab:res_ablation}. Furthermore, it leads to poor ASR performances.

\sstitle{Effects of Sparsity Loss}
Without the sparsity loss, the generated adversarial perturbation is fully dense (\ie sparsity = $100$\,\%). Specifically, when the threat model is Emo18 and the adversarial examples are transferred to attack wav2vec 2.0 and WavLM, the test ASR drops considerably.

\sstitle{Effects of Quantisation Loss}
Without the quantisation loss, we can see that the sparsity increases, especially when the threat model is wav2vec 2.0. 

\subsection{Discussion} 
The proposed STAA-Net, while offering advantages in generating sparse and transferable perturbations efficiently, has several limitations that should be taken into account. Firstly, the selection and fine-tuning of weights for different losses, such as $\lambda_m$, $\lambda_s$, and $\lambda_q$, require careful consideration. Finding the optimal balance between these weights is crucial for achieving desired results. Secondly, although many prior works~\cite{NeekharaHPDMK19, 9822974, 9413467} on adversarial attacks in the audio domain primarily use SNR to evaluate the imperceptibility of the adversarial perturbation, there may be scenarios where stealthiness is more critical than ASR. In such cases, additional evaluation metrics, such as human evaluation, can provide a more comprehensive assessment. Lastly, the availability and quantity of training data for the generator can impact its performance. Insufficient or imbalanced data may limit the generator's ability to learn effectively and generate high-quality adversarial examples. 

\section{Conclusion} 
\label{sec:conclusion}
In conclusion, the field of speech emotion recognition (SER) lacks sufficient research on adversarial attacks, with most existing attacks in the audio domain primarily focusing on $\boldsymbol{l}_2$ or $\boldsymbol{l}_\infty$ norm constraints. To address this gap, we proposed STAA-Net, a generator-based attacker that efficiently generates transferable and sparse adversarial perturbations in an end-to-end manner. We trained the generator using Emo18 and WavLM as threat models and produced adversarial examples in a single forward pass. The generated adversarial examples were then used to attack the considered Zhao19 and wav2vec 2.0 models. Experimental results on the DEMoS and IEMOCAP datasets demonstrated the effectiveness of STAA-Net in achieving a balance between sparsity, speed, imperceptibility, transferability, and attack success rate.

In terms of future research directions, there are several directions that can be explored. Firstly, while this work primarily focused on non-targeted adversarial attacks, it would be valuable to investigate targeted attacks and evaluate the efficiency of the generator in producing transferable and sparse audio adversarial examples that are specifically designed to deceive a particular victim model or class. Secondly, there is room for exploring the applicability and performance in other audio tasks, such as automatic speech recognition. Investigating the effectiveness of the method across different audio domains would help assess its versatility and generalisation capabilities. Thirdly, exploring potential defense mechanisms against such adversarial attacks is an interesting area to investigate, which can contribute to the overall robustness of SER models. Lastly, the exploration of automatic weight determination for different loss components is a worthwhile avenue for future investigation. 




\bibliographystyle{IEEEtran}
\bibliography{bibliography}  

\vfill

\end{document}